\documentclass[lettersize,journal]{IEEEtran}
\usepackage{amsmath,amsfonts}
\usepackage{algorithmic}
\usepackage{algorithm}
\usepackage{array}
\usepackage[caption=false,font=normalsize,labelfont=sf,textfont=sf]{subfig}
\usepackage{textcomp}
\usepackage{stfloats}
\usepackage{url}
\usepackage{verbatim}
\usepackage{graphicx}
\usepackage{array}
\usepackage{booktabs}
\usepackage{multirow}
\usepackage{cite}
\usepackage{bbding}
\usepackage{pifont}% \Checkmark,\XSolid,... (需要和pifont宏包共同使用)
\usepackage{fontawesome}  % \faCheck,\faTimes
\usepackage[colorlinks,
linkcolor=blue,
anchorcolor=blue,
citecolor=blue
]{hyperref}
\newcommand{\cmark}{\ding{51}} % Checkmark symbol
\newcommand{\xmark}{\ding{55}} % Cross symbol

\begin{document}

\markboth{%
  \fontsize{5.3pt}{6pt}\selectfont % 7pt 字号，8pt 行距
  This work has been submitted to the IEEE for possible publication. Copyright may be transferred without notice, after which this version may no longer be accessible.%
}{}

\title{RadioLLM: Introducing Large Language Model into Cognitive Radio via Hybrid Prompt and Token Reprogrammings}

\author{Shuai Chen, Yong Zu, Zhixi Feng,~\IEEEmembership{Member,~IEEE,} Shuyuan Yang,~\IEEEmembership{Senior Member,~IEEE,}, Mengchang Li\\
% , Yue Ma, Jun Liu, Qiukai Pan, Xinlei Zhang, Changjun Sun\\
        % <-this % stops a space
\thanks{This work was supported in part by the National Natural Science Foundation of China (Nos.U22B2018, 62276205, 62171357); Qin chuangyuan"Scientist+Engineer" Team Construction Project of Shaanxi Province Under No.2022KXJ-157; and Shaanxi Province Natural Science Basic Research Program Under No. 2023-JC-YB-560. (\textit{Corresponding author:Zhixi Feng.})}
\thanks{All the authors are with the School of Artificial Intelligence, Xidian University, Xi'an 710071, China. (e-mail: shuai\_chen@stu.xidian.edu.cn, 
	yzuwork@stu.xidian.edu.cn, 
	zxfeng@xidian.edu.cn, 
	syyang@xidian.edu.cn,
        % mengchangli@stu.xidian.edu.cn, mayue@xidian.edu.cn, jun\_liu@stu.xidian.e\\du.cn,
        % panqiukai@stu.xidian.edu.cn,
        % xlzhang1@stu.xidian.edu.cn,
        % sunchangju\\n@stu.xidian.edu.cn
        )}

}



\maketitle

\begin{abstract}
The growing scarcity of spectrum resources and rapid proliferation of wireless devices make efficient radio network management critical. While deep learning-enhanced Cognitive Radio Technology (CRT) provides promising solutions for tasks such as radio signal classification (RSC), denoising, and spectrum allocation, existing DL-based CRT frameworks are typically task-specific and lack scalability in diverse real-world applications. This limitation naturally leads to the exploration of Large Language Models (LLMs), whose exceptional cross-domain generalization capabilities offer new potential for advancing CRT. To bridge this gap, we propose RadioLLM, a novel framework that integrates Hybrid Prompt and Token Reprogramming (HPTR) for combining radio signal features with expert knowledge, and a Frequency-Attuned Fusion (FAF) module for enhanced high-frequency feature modeling. Extensive evaluations on multiple benchmark datasets demonstrate that RadioLLM achieves superior performance compared to existing baselines in the majority of testing scenarios.
\end{abstract}


\section{Introduction}
 \IEEEPARstart{W}{ith} the exponential proliferation of wireless devices and the escalating scarcity of spectrum resources, the efficient management and optimization of constrained wireless network resources has emerged as a pivotal challenge in modern communication systems \cite{newref-1,newref-5}. While the integration of artificial intelligence (AI) with cognitive radio technology (CRT) presents a transformative paradigm for dynamic spectrum sharing and communication quality enhancement \cite{newref-2}, traditional machine learning (ML)-based methods for precise radio cognition are hindered by device-intrinsic noise and an increasing number of interference sources \cite{tii-chen}, rendering them effective only under specific operational conditions. In contrast, deep learning (DL) approaches demonstrate superior performance in radio signal classification (RSC), spectral denoising, and adaptive resource allocation through data-driven optimization frameworks.

\begin{figure}[!t]
\centerline{\includegraphics[width=\columnwidth]{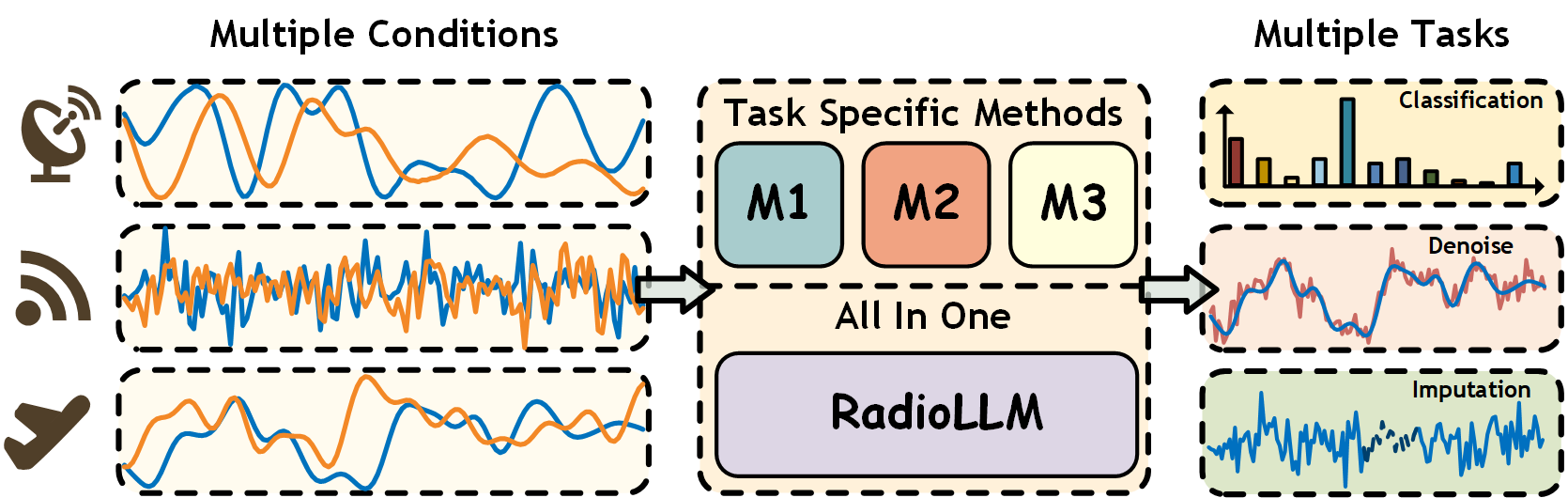}}
	\caption{Comparison of existing CRT frameworks with our proposed approach.}
	\label{fig1}
\end{figure}

Contemporary DL-based CRT architectures, though effective in reducing system complexity via end-to-end training, exhibit inherent task specificity and signal-type dependency. As illustrated in Fig \ref{fig1}, their deployment necessitates extensive task-specific configurations, thereby constraining operational scalability in heterogeneous industrial environments. This limitation underscores the critical need for developing universal CRT frameworks capable of handling cross-domain signal processing task.

% The advent of LLMs has reignited interest in Artificial General Intelligence (AGI). LLMs, pretrained on large-scale datasets using autoregressive techniques, demonstrate generalization capabilities far beyond traditional models, making them promising for CRT\cite{jsac_1}. Despite their potential, LLMs face challenges in CRT applications, such as the need for significant computational resources and architectures tailored to radio signals.

The emergence of LLMs has rekindled significant interest in the pursuit of Artificial General Intelligence (AGI). Pretrained on vast datasets using autoregressive learning techniques, LLMs exhibit generalization capabilities that far surpass those of conventional models, positioning them as promising candidates for application in CRT\cite{jsac_1}. Nevertheless, their deployment in CRT contexts faces several challenges, including substantial computational demands and the need for architectures specifically adapted to the characteristics of radio signals. 

Although LLMs have been employed in applications such as semantic communication \cite{llmref-1}, network optimization \cite{llmref-2}, and spectrum sensing \cite{llmref-3,llmref-13}, existing approaches still exhibit several limitations: 

\textbf{1) Limited Cognitive Understanding of Radio Signals:} Conventional LLMs, trained predominantly on textual data, demonstrate fundamental limitations in semantic comprehension when processing radio signals. Existing approaches in \cite{llmref-3, llmref-14} rely on text descriptions and external tools to transform signals into natural language prompts, which introduces two critical limitations: First, the textualization process inevitably discards essential signal features. Secondly, such methods prevent models from directly learning spatiotemporal patterns from raw signal representations. Building on \cite{llmref-4}, we employ token reprogramming to directly map raw I/Q sequences into LLM-compatible embeddings, thereby eliminating textual intermediaries and enabling true end-to-end signal cognition.

\textbf{2) Inefficient Expert Knowledge Integration:}  
Prompt engineering techniques empower LLMs with task-specific knowledge through carefully designed instructions. However, existing knowledge injection methods  \cite{llmref-4,llmref-7} rely on fixed textual templates that incorporate expert knowledge via dataset background, task descriptions, and statistical information. This approach suffers from two critical limitations: First, excessive irrelevant lexical elements (e.g., prepositions, modifiers) in these templates consume valuable prefix token allocation. Second, the elongated prompt token sequences tend to dilute the feature representation of subsequent signal tokens.

To address these issues, we propose Hybrid Prompting, whose core innovation lies in retrieving the top-K most semantically similar anchors from software-based embedding space to replace conventional hardware-fixed text prompts. Compared with hardware-prompt approaches, our method achieves a 31.85\% improvement in inference speed while maintaining a 0.85\% enhancement in downstream task accuracy.

\textbf{3) Challenges in Capturing High-Frequency Signal Features:}  
The long-range attention mechanisms in LLMs excel at capturing global low-frequency signal patterns \cite{llmref-9, li2025exploring}, yet demonstrate limited sensitivity to high-frequency features such as transient pulses and phase discontinuities \cite{llmref-8, wang2020high}. While existing approach \cite{llmref-13} attempts to enhance perceptual capabilities through task-specific adapter modules, these solutions fail to achieve cross-task transferability and generalization. To overcome these limitations, we propose the Frequency-Attuned Fusion (FAF) module, which synergistically integrates high-frequency local features extracted by CNNs with the global contextual representations learned by LLMs, thereby significantly improving performance in RSC tasks.

In this paper, we introduce RadioLLM, which integrates Hybrid Prompt and Token Reprogramming (HPTR) with CRT using LLMs. HPTR couples expert knowledge with radio signal features, leveraging LLMs' world knowledge for semantic and high-dimensional feature extraction. Additionally, the FAF module improves the modeling of high-frequency information, while a lightweight decoder maps features back to the original signal space. Our contributions include:
\begin{enumerate}
	\item
	We propose a novel multimodal RadioLLM that achieves versatile CRT applications across diverse scenarios through multi-task joint learning. Leveraging the LLM's inherently rich world knowledge, we explore its application in CRT by employing reprogramming techniques. This approach enables the direct processing of radio signals by the LLM, significantly enhancing its cognitive capabilities regarding diverse signal types and reducing reliance on manual prompt engineering.
	\item
	To mitigate computational overhead and memory consumption, we introduce an innovative hybrid prompt technique that combines software and hardware prompts. This method involves identifying the top K semantically similar anchors within a joint semantic space for template text prompt embeddings. These anchors serve as concise and contextually relevant prompts, effectively eliminating unnecessary filler words while maintaining strong correlations with signal features. This streamlined prompting technique optimizes the model's performance in CRT tasks by ensuring that the prompts are both succinct and rich in pertinent information.
	\item 
	We designed the FAF module to enhance the LLM's ability to model high-frequency features by fusing high-frequency and low-frequency information, thereby improving the performance of the downstream classification task while ensuring the performance of the generation task.
\end{enumerate} 
\section{RELATED WORK}
\subsection{Traditional DL-Based CRT Framework}
DL-based methods have been widely applied to critical tasks in CRT, such as RSC, signal denoising, and signal recovery. In RSC, supervised learning (SL) approaches have shown significant progress. For example, PETCGDNN used CNNs and GRUs for feature extraction to build an efficient modulation recognition model \cite{zhang2021efficient}. Additionally, a multi-view fusion RSC method was proposed, which leverages features from the time, frequency, and time-frequency domains to enhance performance \cite{ke2021real}.

The challenge of obtaining high-quality labeled data in wireless communication has driven the adoption of self-supervised learning (SSL). A Transformer-based SSL framework, TCSSAMR, was proposed for RSC \cite{kong2023transformer}, and MCLHN utilized masked contrastive learning with hard negatives to enhance signal diversity \cite{xiao2024mclhn}. For signal denoising and partial recovery, a Deep Denoising Network was proposed using residual learning \cite{kaushal2016better}, and a time-frequency domain autoencoder was developed for denoising \cite{chen2024generative}.

While these methods excel in specific tasks, they are limited to individual applications and specific data types. To address this, we propose the LLM architecture as a universal CRT framework, pioneering AGI applications in CRT.

\subsection{Previous LLM-Based CRT Framework}
The emergence of LLMs has brought significant advancements to AI, influencing domains such as time series analysis and CRT. For example, LLMs have been used in semantic communication systems for contextual understanding \cite{llmref-1} and in 6G edge computing for user association and resource allocation \cite{llmref-2}. However, these applications are mostly limited to telecommunications language understanding.

% WirelessLLM has extended LLM applications to CRT tasks like power allocation and spectrum sensing, advancing AGI integration in CRT \cite{llmref-3}. Yet, it relies on external tools, lacks an end-to-end signal processing pipeline, and overlooks critical tasks like WSC and signal denoising.

% To overcome these challenges, we build on recent advances in time-series LLM research \cite{llmref-4}, \cite{llmref-7}, \cite{llmref-5} by incorporating reprogramming and hybrid prompting techniques. This enables the development of a unified and versatile CRT framework capable of handling diverse signal types and tasks.

While WirelessLLM extends LLMs to CRT tasks such as power allocation and spectrum sensing \cite{llmref-3}, it depends on external tools, lacks an end-to-end pipeline, and misses key capabilities like WSC and signal denoising. Similarly, LLM4WM, a multitask framework for channel modeling \cite{llmref-13}, employs task-specific adapters that impede the learning of unified representation and model coherence. 

To address these limitations, we leverage recent advances in time-series LLMs \cite{llmref-4}, \cite{llmref-7}, \cite{llmref-5} through model reprogramming and hybrid prompting, enabling a unified and versatile CRT framework for diverse signals and tasks.

\section{Methodology}
\begin{figure*}[!t]%%图 指俩栏排版中图占两栏，不加则适应双栏!htp取消美学标准,美学标准很烂 h当前位置 t顶部 b底部 p浮动页 <216mm  \linewidth
	\centerline{\includegraphics[width=\textwidth]{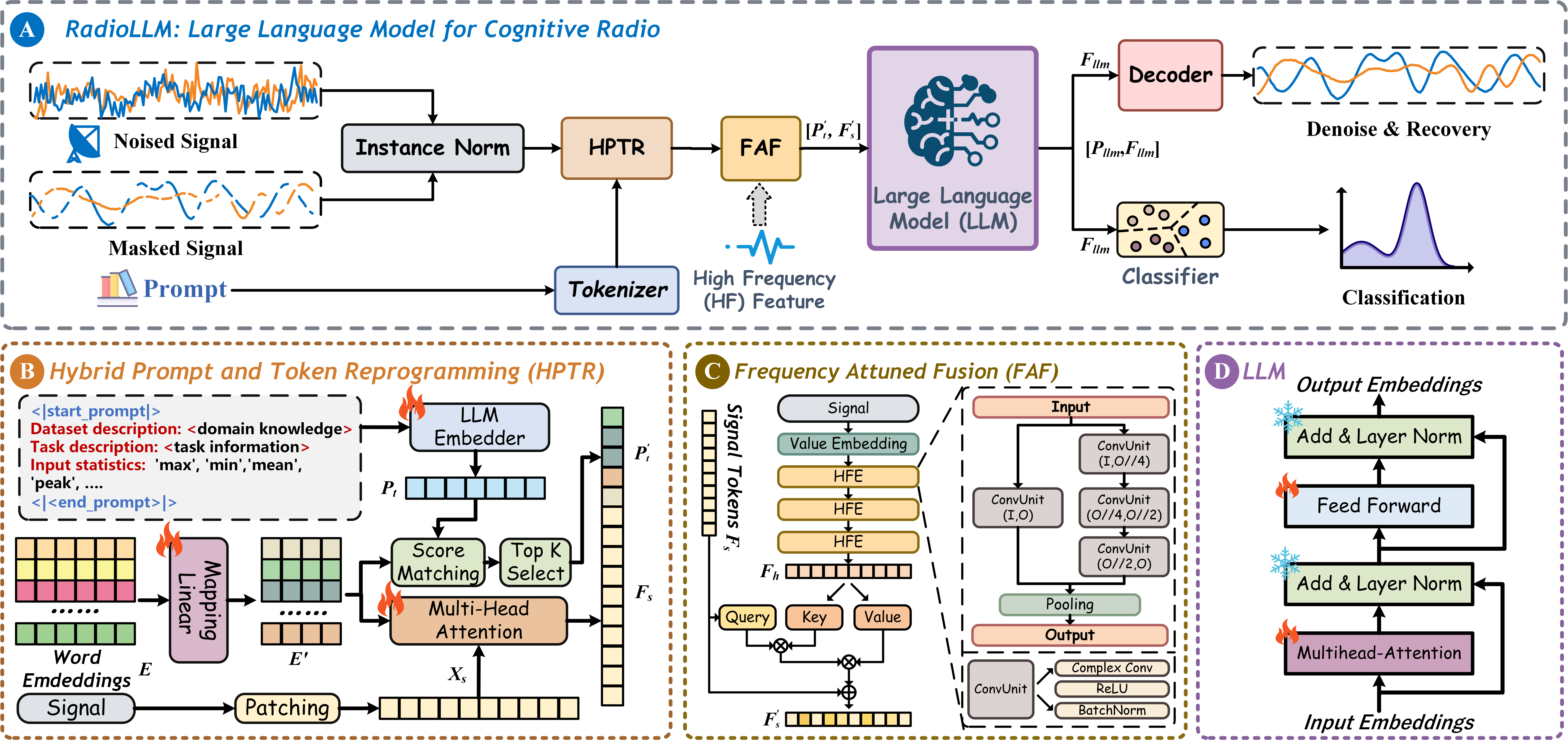}}
	\caption{The model framework of RadioLLM. The input radio signal is preprocessed to generate signal embeddings \( X_{s} \) (A). In the HPTR stage (B), \( X_{s} \) is reprogrammed with semantic anchors \(E'\), and top-K semantic anchors are selected as prefix prompt \( P'_t\). The FAF stage (C) injects high-frequency features to address the transformer's low-pass filtering tendency. Finally, the enhanced embeddings and prefix prompts are fed into the LLM (A\&D), which outputs denoised signals \( O_s \) or classification results depending on the task.}
	\label{fig2}
\end{figure*}

% \textbf{Overview}: As illustrated in Figure \ref{fig2}, RadioLLM consists of two key components. Given an input radio signal, we first partition it into patches to obtain the signal embedding. Next, expert knowledge prompts are combined with the top-K semantically similar anchors retrieved from word token embeddings, enriched with extensive world knowledge, to serve as prefix prompts for the signal embedding. Finally, high-frequency features extracted by a CNN are nonlinearly coupled with the signal’s low-frequency information, and the fused features are fed into the pre-trained LLM.
\textbf{Overview}: As shown in Fig.~\ref{fig2}, RadioLLM is designed with two primary components. First, the input radio signal is divided into patches to generate its signal embedding. The first component involves combining expert knowledge prompts with the top-K semantically similar anchors, retrieved from word token embeddings enriched with extensive external knowledge. These anchors and prompts are used to construct prefix prompts for the signal embedding. The second component integrates high-frequency CNN features with low-frequency signal information, feeding the fused output into the pre-trained LLM.

In this work, GPT-2 is adopted as the backbone. During training, we not only learn the mapping function between inputs and outputs but also fine-tune GPT-2 using the LoRA technique \cite{llmref-12}.
\subsection{Problem Statement}
In cognitive radio systems, the signal received by the secondary user device is often distorted due to the effects of the wireless channel and environmental noise. The received IQ signal \( r(t) \in \mathbb{R}^{2 \times L} \) can be modeled as:
\begin{equation}
	r(t) = h(t) * s(t) + n(t),
	\label{eq1}
\end{equation}
where \( s(t) \) is the transmitted signal, \( h(t) \) represents the channel response (e.g., path loss, multipath effects), and \( n(t) \) is the additive white Gaussian noise (AWGN). Here, \( L \) is the signal length, and 2 corresponds to the in-phase (I) and quadrature (Q) components.

The goal is to design a unified framework that can address multiple downstream tasks, including denoising, signal recovery, and classification. Let \( T \) denote the task-specific target:
\begin{itemize}
	\item For signal reconstruction tasks (including denoising and recovery), \( T = \hat{s}(t) \), where \( \hat{s}(t) \) represents the reconstructed signal or a task-specific attribute of the original transmitted signal.
	\item For classification tasks, \( T = y \), where \( y \in \{1, 2, \dots, C\} \) is the class label associated with the signal.
\end{itemize}

To address these tasks, we aim to learn a unified model \( F(r(t); \Theta) \), parameterized by \( \Theta \), which maps the received signal \( r(t) \) to the target \( T \). The unified optimization objective can be expressed as:
\begin{equation}
	\hat{\Theta} = \arg\min_{\Theta} \mathbb{E}[\mathcal{L}(F(r(t); \Theta), T)],
	\label{eq2}
\end{equation}
where \( \mathcal{L}(\cdot, \cdot) \) is a task-specific loss function.
\subsection{Hybrid Prompt and Token Reprogramming}
Hardware prompts are widely used to inject expert knowledge into LLMs but often involve verbose, template-based structures that dilute meaningful information and increase computational cost. In contrast, pretrained software prompts encode general world knowledge but lack domain-specific relevance. To address these limitations, we propose a hybrid prompt mechanism, which combines hardware prompts with a reduced set of software prompts to achieve efficient and domain-relevant representations.

Given pretrained word token embeddings \( E \in \mathbb{R}^{V \times D} \), where \( V \) is the vocabulary size and \( D \) is the embedding dimension, we derive a reduced set of semantic anchors \( E' \in \mathbb{R}^{V' \times D} \) (\( V' \ll V \)) via a mapping function \( f(E) \). For a hardware prompt \( T \), tokenized and embedded as \( P_t \in \mathbb{R}^{L_T \times D} \), we compute a hybrid prompt \( P'_t \in \mathbb{R}^{K \times D} \) by selecting the top-\( K \) most similar embeddings from \( E' \), formalized as:
\begin{equation}
	P'_t[k, :] = E'\left[\text{ArgTopK}\left(\max_{i=1}^{L_T} \gamma(P_t[i, :], E'), K\right), :\right],
	\label{eq3}
\end{equation}
where \( \gamma(\cdot) \) is the cosine similarity function defined as:
\begin{equation}
	\gamma(P_t[i, :], E') = \frac{P_t[i, :] \cdot E'^T}{\|P_t[i, :]\| \cdot \|E'\|}.
	\label{eq4}
\end{equation}

Here, \( P'_t \in \mathbb{R}^{K \times D} \) represents the hybrid prompt constructed by selecting the top-\( K \) embeddings from \( E' \), based on the maximum cosine similarity scores over all \( L_T \) tokens in \( P_t \).

Since LLMs are trained on textual tokens, radio signals cannot be directly understood by LLMs nor described losslessly in natural language. Therefore, it is necessary to reprogram the radio signal sequences into semantic tokens interpretable by the LLM. To achieve this, we leverage a multi-head cross-attention layer for the reprogramming process. Specifically, we use \( X_s \) as the query matrix and \( E' \) as the key and value matrices. The reprogramming operation for each attention head is defined as follows:
\begin{equation}
	\text{Attention}(X_s, E', E') = \text{softmax}\left(\frac{X_s W_q (E' W_k)^T}{\sqrt{d_k}}\right) E' W_v,
	\label{eq5}
\end{equation}
where \( W_q \), \( W_k \), and \( W_v \) are the learnable projection matrices for the query, key, and value, respectively, and \( d_k \) is the dimension of the key vectors. The outputs from all attention heads are concatenated and passed through a linear transformation. This linear projection maps the output of the attention layer to the LLM-compatible dimension, resulting in the signal tokens \( F_s \in \mathbb{R}^{P \times D} \), where \( P \) is the number of patches and \( D \) is the feature dimension. These signal tokens are then fed into the LLM for further processing.
\subsection{Frequency Attuned Fusion}
Existing LLMs, based on the Transformer architecture, excel at capturing low-frequency global information through the attention mechanism but are less sensitive to high-frequency features. In contrast, CNNs naturally excel at modeling high-frequency information. Based on this understanding, we propose the FAF module to augment the signal tokens \( F_s \), enhancing sensitivity to high-frequency information in radio signals.

The FAF module consists of three high-frequency extraction (HFE) layers, with the structure of each HFE layer shown in Fig.~\ref{fig2} (C). Each HFE layer leverages convolution to detect local variations, ReLU to enhance non-linear features, and pooling to compress redundant information, effectively extracting high-frequency features from the input data. This design enables the network to better capture fine-grained details in the input. 

The FAF module takes the raw signal \( r(t) \) as input and outputs high-frequency features \( F_h \in \mathbb{R}^{P \times D} \) after three HFE layers. Subsequently, original signal tokens are reprogrammed with the high-frequency features \( F_h \), while also being incorporated as a supplementary component. This process yields the frequency-enhanced signal tokens \( F_s' \in \mathbb{R}^{P \times D} \). By integrating both global low-frequency information and fine-grained high-frequency details, these enhanced tokens enable the model to achieve a more comprehensive representation of the input radio signal.
\subsection{Output Projection}
The features \( F_s' \) and \( P_t' \) are fed into the fine-tuned LLM module (Fig.~\ref{fig2} (D)) to obtain \( F_{\text{llm}} \) and \( P_{\text{llm}} \). After discarding the prefix \( P_{\text{llm}} \), \( F_{\text{llm}} \) is passed to the decoder to generate the output \( O_s \in \mathbb{R}^{2 \times L} \). In this work, we explore two decoding strategies: a linear layer for direct mapping and a shallow Transformer decoder that utilizes self-attention to capture complex dependencies, resulting in improved reconstruction. The pretraining objective is to minimize the mean squared error (MSE) loss between \( O_s \) and the ground truth \( s'(t) \).
\subsection{Training Strategy}
The overall pretraining pipeline of RadioLLM and its application to downstream tasks are summarized in Algorithm \ref{alg:radioLLM}. The process begins with the configuration of hyperparameters, including the network structure, optimizer, dataset parameters, and early stopping criteria. During pretraining, most parameters of the LLM remain frozen, with only a subset updated via LoRA \cite{llmref-2}. To ensure balanced learning when processing multiple datasets within the same epoch, a loss balancing factor $ b_i $ is applied to normalize the gradient contributions from each dataset, which helps mitigate biases caused by differences in data scale or complexity. Upon completion of pretraining, only the non-frozen parameters are retained to reduce storage overhead.

% For downstream tasks, \( F'_s \), extracted from the LLM output, is used as the feature representation for classification tasks. After pooling, it is fed into a linear classification head. The decoder's output \( O_s \) is directly used as the prediction for denoising and completing tasks. 

For downstream tasks, two task-specific adaptation strategies are employed. In classification tasks, a feature representation $ F'_s $ is extracted from the LLM output, followed by a pooling operation, and then fed into a linear classification head. For denoising or signal reconstruction tasks, the decoder output $ O_s $ is directly used as the final prediction.

\begin{algorithm}[!h]
    \caption{RadioLLM Pretraining and Applications}
    \label{alg:radioLLM}
    \textbf{Input}: Unlabeled dataset \( \mathcal{D}^{U} = \{D^{U}_{i}\}_{i=1}^N \), labeled dataset \( \mathcal{D} = \{D_{i}\}_{i=1}^N \), balancing factors \( \{b_i\}_{i=1}^N \), model parameters \( \Theta \), hyperparameters (learning rate \( lr \), batch size \( B \), total epochs \( E \).\\
    \textbf{Output}: Pretrained model parameters \( \Theta^{*} \), recovery signal \( O_s \), classification result \( Y_s \).

    \begin{algorithmic}[1] %[1] enables line numbers
        \STATE \textbf{Stage 1: Pretraining RadioLLM}
        \STATE \textbf{Initialize} the parameters of \( \Theta \).
        \FOR{\( e \leftarrow 1 \) to \( E \) epochs}
            
            \FOR{\( \text{batch} \ \{r(t)\}_{i=1}^B \in D^{U}_{i} \)}
                \STATE \( F_s, P'_t \leftarrow F_{HPTR}(\{r(t)\}_{i=1}^B) \) 
                \STATE \( F'_s \leftarrow F_{FAF}(F_s) \) 
                \STATE \( [P_{llm}, F_{llm}] \leftarrow F_{LLM}(F'_s, P'_t) \) 
                \STATE \( O_s \leftarrow F_{Decoder}(F_{llm}) \) 
                \STATE \( \mathcal{L} \leftarrow b_i \cdot \mathcal{L}_i(O_s, T) \) \hfill \(\triangleright\) Compute MSE loss
                \STATE \( \Theta \leftarrow \Theta - lr_{ssl} \nabla_{\Theta} \mathcal{L} \) 
            \ENDFOR
         
            \IF{\( \mathcal{L} \) does not decrease for 20 consecutive epochs}
                \STATE \textbf{break}
            \ENDIF
        \ENDFOR
        \STATE \( \Theta^{*}  \leftarrow \Theta \)
        \STATE Save the non-frozen parameters of \( \Theta^{*} \).
        \STATE \textbf{Stage 2: Downstream Tasks Application}
    \IF{\text{Task} = \text{RSC}}
        \STATE Initialize the parameters of classifier \(F_{fc}( \cdot, \Theta_{fc}).\)
        \FOR{\( \text{batch} \ \{r(t), y\}_{i=1}^B \in D_i \)}
            \STATE \( Y_s \leftarrow F_{fc}(\text{Pool}(F'_s), \Theta_{fc}) \)
            \STATE \( \mathcal{L} \leftarrow\mathcal{L}_i(Y_s, y) \) \hfill \(\triangleright\) Compute Cross-Entropy loss
            \STATE \( \Theta_{fc} \leftarrow \Theta_{fc} - lr_{rsc} \nabla_{\Theta_{fc}} \mathcal{L} \) 
        \ENDFOR
    \ELSE
        \STATE \( \text{Output} \leftarrow O_s \) 
    \ENDIF
    \end{algorithmic}
\end{algorithm}
\section{Experiments}
\subsection{Data Overview}
In this study, seven publicly available datasets were utilized to develop and evaluate the proposed method, including RadioML2016.10a (RML16A), RadioML2016.10b (RML16B) \cite{rml2016ab}, RadioML2016.04c (RML16C) \cite{rmlc}, RadioML2022 (RML22) \cite{rml2022}, RadioML2018.01a (RML18A) \cite{rml2018}, the ADS-B dataset \cite{adsb}, and the Wi-Fi dataset \cite{wi-fi}. To ensure fair evaluation and prevent data leakage in downstream tasks, all datasets were partitioned into training, validation, and test sets in an 8:1:1 ratio.

For pretraining, we utilized a combination of the RML16A, RML16B, RML16C, RML18A, ADS-B, and WiFi datasets. Specifically, only high-quality samples with (\(\text{SNR} \geq 14\,\mathrm{dB}\)) were selected from the RML series datasets to ensure robust feature learning from clean signals. For the ADS-B dataset, the entire training set was used without SNR filtering to capture the full range of real-world signal variations. For the WiFi dataset, 5\% of the training samples were randomly chosen to provide representative diversity while maintaining computational efficiency. This pretraining strategy enabled the model to leverage both high-quality and diverse real-world data, thereby enhancing its generalization capability.

\begin{figure}[h!]
\centering
\includegraphics[width=0.48\textwidth]{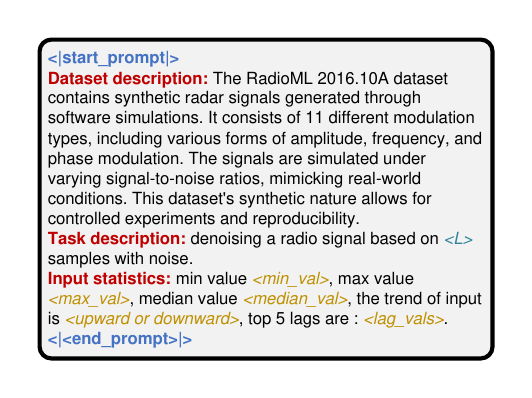}
\caption{An illustrative example of the prompt template used in this study. The template incorporates domain-specific descriptions, statistical attributes, and task-specific instructions, which are color-coded for clarity. This unified prompt design enables the model to process heterogeneous signal information with rich contextual and structural cues.}
\label{fig:prompt-template}
\end{figure}

To fully exploit the unique characteristics of each dataset and enhance the model’s adaptability to diverse downstream tasks, we developed a modular prompt design framework. An illustrative example of our prompt template is shown in Fig.~\ref{fig:prompt-template}, where the prompt is composed of three main components: a dataset description, a task description, and input statistics.

Specifically, for each dataset, we constructed a tailored dataset description that encapsulates its domain attributes, such as signal source, modulation types, and SNR conditions. In addition, for each pretraining objective—including denoising, recovering, and classification—we designed corresponding task-specific prompts that explicitly define the nature of the learning problem. Furthermore, in order to provide the model with fine-grained contextual cues, we incorporated input statistics into the prompt structure. These statistical descriptors (e.g., minimum, maximum, and median values, overall trend, and top lag features) are dynamically generated based on each individual data instance.

This unified and extensible prompt design enables the model to seamlessly process heterogeneous datasets and diverse learning tasks, while leveraging contextual information at both global and instance levels.

% To fully exploit the unique characteristics of each dataset and enhance the model’s adaptability to diverse downstream tasks, we developed a modular prompt design framework. Specifically, for each dataset, we constructed a tailored dataset description that encapsulates its domain attributes, such as signal source, modulation types, and SNR conditions. In addition, for each pretraining objective, including denoising, recovering, and classification. We designed corresponding task-specific prompts that explicitly define the nature of the learning problem. Furthermore, in order to provide the model with fine-grained contextual cues, we incorporated input statistics into the prompt structure. These statistical descriptors (e.g., minimum, maximum, and median values, overall trend, and top lag features) are dynamically generated based on each individual data instance. An illustrative example of our prompt template is shown in Fig.~\ref{fig:prompt-template}, where the prompt is composed of a dataset description, task description, and input statistics.

% This unified and extensible prompt design enables the model to seamlessly process heterogeneous datasets and diverse learning tasks, while leveraging contextual information at both global and instance levels.
\subsection{Dataset Details}
\textbf{RML16A}: The RadioML2016.10a dataset comprises 220,000 modulated signals generated under simulated interference conditions, including carrier frequency offset and additive white Gaussian noise (AWGN). Each signal consists of 128 complex-valued I and Q samples. The dataset covers three analog modulation types—Wideband Frequency Modulation (WBFM), Amplitude Modulation Single Sideband (AM-SSB), Amplitude Modulation Double Sideband (AM-DSB)—and eight digital modulation types: Continuous Phase Frequency Shift Keying (CPFSK), Binary Phase Shift Keying (BPSK), 8-Phase Shift Keying (8PSK), Gaussian Frequency Shift Keying (GFSK), 16-Quadrature Amplitude Modulation (16QAM), 64-Quadrature Amplitude Modulation (64QAM), 4-Pulse Amplitude Modulation (4PAM), and Quadrature Phase Shift Keying (QPSK).

\textbf{RML16B}: The RadioML2016.10b dataset contains approximately 1.2 million modulated signals, covering the same modulation categories as RML16A except for AM-SSB. Each signal sample also consists of 128 I/Q points.

\textbf{RML16C}: The RadioML2016.04c dataset includes 162,060 modulated signals with the same modulation types as RML16A. Each sample similarly contains 128 I/Q data points.

\textbf{RML22}: The RadioML2022 dataset comprises 420,000 modulated signal samples, including eight digital modulation types—8PSK, BPSK, CPFSK, GFSK, 4PAM, 16QAM, 64QAM, and QPSK—and two analog modulation types: AM-DSB and WBFM. The signals span an SNR range from \(-20\,\mathrm{dB}\) to \(20\,\mathrm{dB}\), with a step size of \(2\,\mathrm{dB}\).

\textbf{RML18A}: The RadioML2018.01a dataset serves as a large-scale benchmark for signal modulation classification, containing 2,555,904 signal samples across 24 modulation types (e.g., OOK, BPSK, QPSK, 16QAM) and 26 SNR levels from \(-20\,\mathrm{dB}\) to \(30\,\mathrm{dB}\) in \(2\,\mathrm{dB}\) increments. Each sample consists of 1024 I/Q points, with 4096 samples per modulation type per SNR level, offering a balanced and comprehensive dataset.

\textbf{ADS-B}: The ADS-B dataset consists of 63,105 Automatic Dependent Surveillance–Broadcast signals captured in real-world open environments. Each signal is sampled at \(50\,\mathrm{MHz}\) and contains 3000 sampling points, spanning 198 distinct signal classes.

\textbf{Wi-Fi}: The Wi-Fi dataset, described in \cite{wi-fi}, consists of over-the-air transmissions captured using 16 USRP X310 devices with a transmitter–receiver distance of 2 feet. The raw signals were processed into sequences of 128 complex-valued I/Q samples to maintain consistency with other datasets.

\subsection{Implementation Details}
\label{Implementation}
During the training process, we conducted all experiments, including pretraining, comparisons, and ablations, on two A800 Ubuntu servers using the PyTorch 2.3.0 framework. The AdamW optimizer (weight decay = $5 \times 10^{-3}$) was employed with an initial learning rate of $5 \times 10^{-5}$. A linear warmup and decay schedule was applied during the pretraining phase, with the warmup phase covering 10\% of the total training epochs. The training process was capped at 50 epochs, and the learning rate was halved if the validation loss stagnated for 5 consecutive epochs. Additionally, early stopping was triggered if validation loss did not improve for 20 consecutive epochs.

For downstream tasks like classification and denoising, a cosine annealing learning rate schedule was employed, starting with an initial learning rate of $5 \times 10^{-5}$. All experiments were conducted under uniform settings to maintain fairness across methods and tasks. Inference time was measured in seconds per batch, with the batch size fixed at 128.
\subsection{Data Augmentation}
\subsubsection{Phase Rotation}
Phase rotation is performed by rotating the phase of the original signal. The process can be mathematically described as:
\begin{equation}
s'(t) = s(t) \cdot e^{j\theta},
\end{equation}
where \(s(t)\) is the original signal, \(\theta\) is the rotation angle, \(j = \sqrt{-1}\), and \(s'(t)\) is the phase-rotated signal. The angle \(\theta\) is typically chosen randomly from the interval \([0, 2\pi)\).
\subsubsection{Signal Reverse}
Signal reverse enhances the signal by flipping it along the time axis. The process is defined as:
\begin{equation}
s'(t) = s(-t),
\end{equation}
where \(s(t)\) is the original signal, and \(s'(t)\) is the reversed signal. For discrete signals, the operation is expressed as:
\begin{equation}
s'[n] = s[N-1-n],
\end{equation}
where \(N\) is the total length of the signal, and \(n\) is the discrete time index.
\subsubsection{Time Warp}
Time warping is a method to enhance signals by altering their temporal dynamics through nonlinear time axis transformations. The process is mathematically described as:
\begin{equation}
s'(t) = s(\phi(t)),
\end{equation}
where \(s(t)\) is the original signal, \(\phi(t)\) is the time warping function, and \(s'(t)\) is the time-warped signal. The warping function \(\phi(t)\) must satisfy:
\begin{itemize}
    \item \(\phi'(t) > 0\) (monotonicity),
    \item \(\phi(t) \in [0, T]\), where \(T\) is the total duration of the signal.
\end{itemize}

For discrete signals, the time warping operation is expressed as:
\begin{equation}
s'[n] = s[\phi(n)],
\end{equation}
where \(\phi(n)\) is the discrete time warping function.

\subsection{Evaluation Metrics}
To meet practical requirements, we employed widely used metrics for quantitative assessment, including Overall Accuracy (OA), Cohen’s Kappa coefficient (Kappa), and Structural Similarity Index Measure (SSIM). OA and Kappa were utilized to evaluate classification performance, measuring overall accuracy and agreement beyond chance, respectively. SSIM was used to assess the structural similarity between predicted and ground truth signals, providing a perceptually aligned evaluation of reconstruction quality.
\subsubsection{OA}
OA is a widely used metric in classification tasks to measure the proportion of correctly classified samples. It is defined as:

\[
\text{OA} = \frac{\sum_{i=1}^k n_{ii}}{N},
\]
where $n_{ii}$ is the number of correctly classified samples for class $i$, $k$ is the total number of classes, $N$ is the total number of samples.

\subsubsection{Kappa Score}

Kappa is a statistical measure of inter-rater agreement or classification reliability. It accounts for the possibility of agreement occurring by chance. $\text{Kappa}$ is defined as:

\[
\text{Kappa} = \frac{\text{OA} - \text{PE}}{1 - \text{PE}},
\]
where $\text{PE}$ is the expected agreement by chance, calculated as:
  \[
  \text{PE} = \sum_{i=1}^k \left( \frac{\sum_{j=1}^k n_{ij} \cdot \sum_{j=1}^k n_{ji}}{N^2} \right),
  \]
where $n_{ij}$ is the number of samples classified as class $j$ when their true label is class $i$. $\text{Kappa}\in[-1,1]$ where 1 indicates perfect agreement, 0 represents chance-level agreement, and negative values suggest systematic disagreement.
\subsubsection{SSIM}

SSIM is a perceptual metric used to measure the similarity between two images. It is particularly designed to evaluate the quality of denoised or reconstructed images by comparing them with their original counterparts. 

The SSIM between two signals, $x$ and $y$, is defined as:

\[
\text{SSIM}(x, y) = \frac{(2\mu_x \mu_y + C_1)(2\sigma_{xy} + C_2)}{(\mu_x^2 + \mu_y^2 + C_1)(\sigma_x^2 + \sigma_y^2 + C_2)},
\]
where $\mu$ denotes mean, $\sigma^2$ variance, $\sigma_{xy}$ covariance, with $C_1=(K_1L)^2$ and $C_2=(K_2L)^2$ as stabilization constants ($L$: value range, $K_1,K_2\ll1$). SSIM approaches 1 for identical signals.

These metrics provide a comprehensive evaluation for both the denoising and classification tasks, ensuring the reliability and quality of the results.
\subsection{Comparison with RSC methods}
\begin{table*}[!t]
\centering
\caption{Comparison of Methods on Classification Tasks. \textbf{BOLD} indicates the best performance, and \underline{UNDERLINED} indicates the second-best performance.}
\label{tab:classification_results}
\setlength{\tabcolsep}{3.6pt} % Reduce column separation
 % Reduce font size
\begin{tabular}{@{}c|cc|cc|cc|cc|cc|cc|cc@{}}
\toprule
\multirow{2}{*}{\textbf{Methods}} 
& \multicolumn{2}{c|}{\textbf{RML16A}} 
& \multicolumn{2}{c|}{\textbf{RML16B}} 
& \multicolumn{2}{c|}{\textbf{RML16C}} 
& \multicolumn{2}{c|}{\textbf{RML22}} 
& \multicolumn{2}{c|}{\textbf{RML18A}} 
& \multicolumn{2}{c|}{\textbf{Wi-Fi}} 
& \multicolumn{2}{c}{\textbf{ADS-B}} \\ \cmidrule(lr){2-3}\cmidrule(lr){4-5}\cmidrule(lr){6-7}\cmidrule(lr){8-9}\cmidrule(lr){10-11}\cmidrule(lr){12-13}\cmidrule(lr){14-15}
& \textbf{OA(\%)} & \textbf{Kappa} 
& \textbf{OA(\%)} & \textbf{Kappa} 
& \textbf{OA(\%)} & \textbf{Kappa} 
& \textbf{OA(\%)} & \textbf{Kappa} 
& \textbf{OA(\%)} & \textbf{Kappa} 
& \textbf{OA(\%)} & \textbf{Kappa} 
& \textbf{OA(\%)} & \textbf{Kappa} \\ \midrule
\textbf{HCGDNN} \cite{HCGDNN} & 54.30 & 0.4973 & 55.30 & 0.5033 & 65.02 & 0.6097 & 56.93 & 0.5215 & 31.29 & 0.2831 & 15.67 & 0.1004 & 44.66 & 0.4434 \\
\textbf{PETCGDNN} \cite{PETCGDNN} & 51.09 & 0.4620 & 50.91 & 0.4546 & 62.12 & 0.5771 & 56.05 & 0.5116 & 28.70 & 0.2560 & \underline{34.59} & \underline{0.3023} & 97.47 & 0.9746 \\
\textbf{MCLDNN} \cite{MCLDNN} & 52.84 & 0.4812 & 52.45 & 0.4716 & 65.63 & 0.6130 & 53.28 & 0.4809 & 42.05 & 0.3953 & 19.41 & 0.1403 & 65.12 & 0.6493 \\
\textbf{CVCNN} \cite{CVCNN} & 54.15 & 0.4957 & 54.06 & 0.4896 & 65.49 & 0.6142 & 54.57 & 0.4953 & 34.79 & 0.3196 & 11.44 & 0.0553 & 97.49 & 0.9747 \\
\textbf{SFS-SEI} \cite{SFS-SEI} & 54.13 & 0.4954 & 54.17 & 0.4907 & 65.45 & 0.6143 & 53.49 & 0.4832 & 36.96 & 0.3422 & 11.45 & 0.0555 & \underline{97.50} & \underline{0.9749} \\
\textbf{ICAMCNET} \cite{ICMACNET} & 53.55 & 0.4891 & 53.43 & 0.4825 & 63.17 & 0.5892 & 54.23 & 0.4914 & 36.50 & 0.3374 & 11.34 & 0.0543 & \textbf{99.09} & \textbf{0.9908} \\
\textbf{CVSRN} \cite{CVSRN} & 47.56 & 0.4232 & 51.50 & 0.4611 & 64.41 & 0.6021 & 50.09 & 0.4454 & 35.74 & 0.3295 & 11.72 & 0.0583 & 94.55 & 0.9452 \\
\textbf{TcssAMR} \cite{tcssamr} & \underline{56.53} & \underline{0.5218} & 55.98 & 0.5109 & \underline{66.21} & \underline{0.6224} & \underline{58.17} & \underline{0.5353} & 47.35 & 0.4506 & 32.86 & 0.2873 & 61.89 & 0.6177 \\
\textbf{SemiAMC} \cite{Semiamc} & 55.32 & 0.5076 & \underline{56.17} & \underline{0.5129} & 64.62 & 0.6056 & 58.08 & 0.5342 & \underline{47.66} & \underline{0.4539} & 29.36 & 0.2502 & 70.42 & 0.7034 \\
\textbf{RadioLLM} & \textbf{58.10} & \textbf{0.5391} & \textbf{58.35} & \textbf{0.5372} & \textbf{68.19} & \textbf{0.6441} & \textbf{59.39} & \textbf{0.5488} & \textbf{52.03} & \textbf{0.4994} & \textbf{35.41} & \textbf{0.3110} & 90.58 & 0.9054 \\
\bottomrule
\end{tabular}
\end{table*}

We conducted extensive performance evaluations of our proposed model, RadioLLM, against several representative RSC approaches, including SL methods such as HCGDNN \cite{HCGDNN}, PETCGDNN \cite{PETCGDNN}, MCLDNN \cite{MCLDNN}, ICMACNET \cite{ICMACNET}, CVCNN \cite{CVCNN}, SFS-SEI \cite{SFS-SEI}, CVSRN \cite{CVSRN}, and SSL methods such as SemiAMC \cite{Semiamc} and TCSSAMR \cite{tcssamr}. For methods without publicly available implementations, we carefully reconstructed network architectures based on original descriptions and retrained them using identical experimental conditions for fair comparisons. All experiments utilized multiple well-established datasets, specifically RML16A, RML16B, RML16C, RML22, RML18A, Wi-Fi, and ADS-B, under the 100-shot scenario, except for ADS-B where 10\% of the data was used due to its limited size.

Table~\ref{tab:classification_results} summarizes the comprehensive performance metrics, including OA and Kappa coefficients for each method across all evaluated datasets under the 100-shot scenario. Clearly, RadioLLM demonstrates state-of-the-art performance on most datasets. Specifically, RadioLLM achieves the highest accuracy of 58.10\%, 58.35\%, 68.19\%, 59.39\%, 52.03\%, and 35.41\% on RML16A, RML16B, RML16C, RML22, RML18A, and Wi-Fi datasets, respectively. This substantial improvement highlights its robust ability to effectively leverage pre-trained knowledge for optimized feature extraction and transfer learning across different signal types.

\begin{figure*}[!h]
    \centering
    \includegraphics[width=\linewidth]{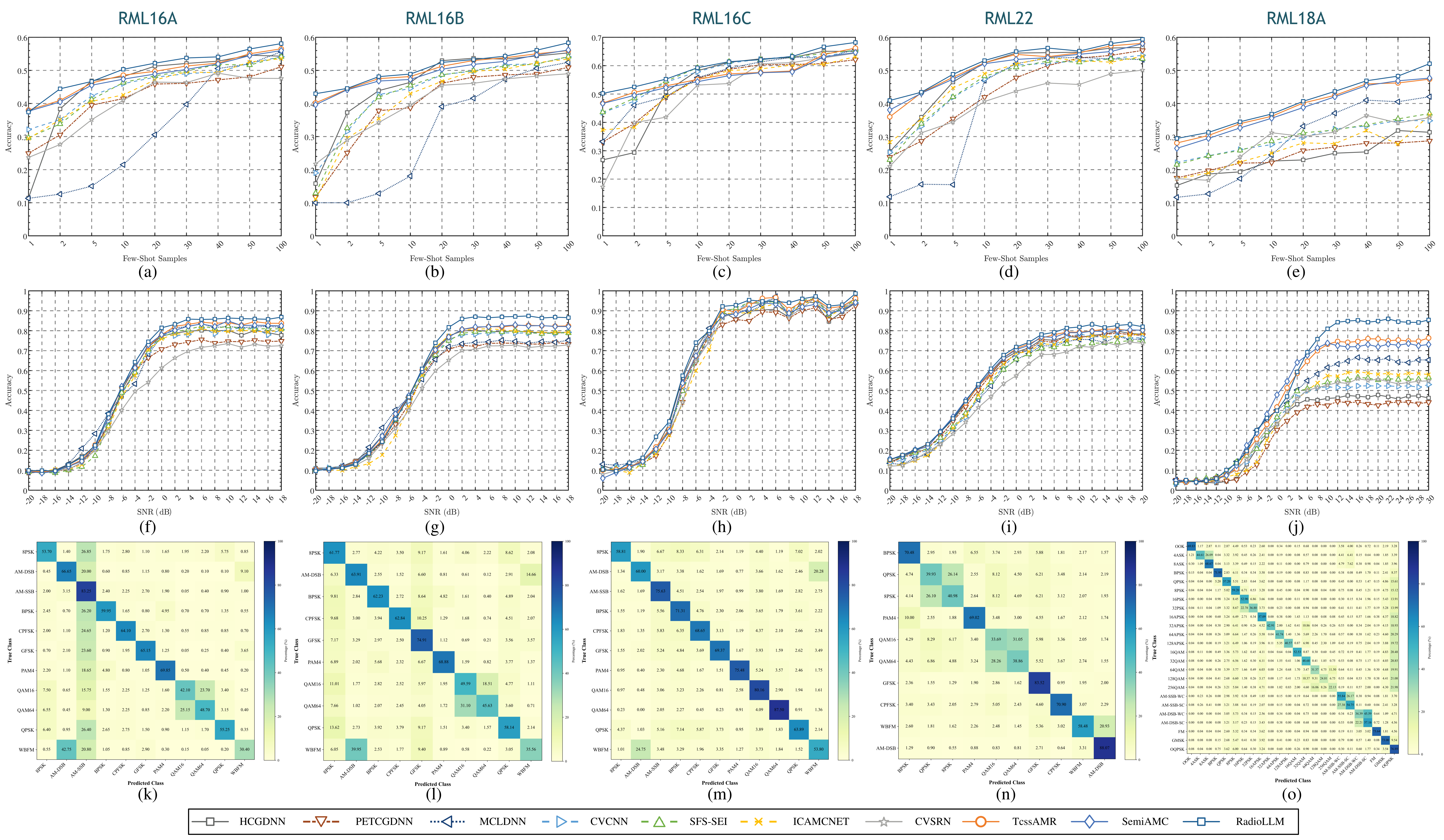}
    \caption{Performance evaluation of RadioLLM across multiple datasets and SNR levels. Each column corresponds to one dataset: RML2016a, RML16B, RML16C, RML22, and RML18A.
(a)–(e) show the OA of RadioLLM compared to other models under varying few-shot sample settings.
(f)–(j) depict the OA under different SNR conditions with the few-shot number fixed at 100.
(k)–(o) present the confusion matrices of RadioLLM on the corresponding datasets, illustrating detailed classification results across modulation classes.}
 \label{fig:acc-snr}
\end{figure*}
% As quantitatively demonstrated in Table~\ref{tab:classification_results}, which presents the OA and Kappa metrics across all evaluated datasets, our proposed RadioLLM framework establishes new state-of-the-art performance benchmarks. The experimental results reveal consistent superiority, with RadioLLM achieving maximum accuracy scores of 58.10\%, 58.35\%, 68.19\%, 59.39\%, 52.03\%, and 35.41\% on the RML16A, RML16B, RML16C, RML22, RML18A and Wi-Fi datasets, respectively. This substantial improvement highlights its robust ability to effectively leverage pre-trained knowledge for optimized feature extraction and transfer learning across different signal types.
On the RML16A dataset, the proposed model outperformed TcssAMR under the 100-shot setting, underscoring the benefits of pre-training for effective feature extraction. However, as illustrated in Fig.~\ref{fig:acc-snr} (k), the model encounters difficulties in distinguishing between similar modulation types, such as 16-QAM and 64-QAM, as well as noise-sensitive signals like WBFM and AM-DSB. Notably, a substantial proportion of samples from several classes are misclassified as AM-SSB, with QPSK exhibiting a 26.4\% misclassification rate into this category.

Furthermore, Fig.~\ref{fig:acc-snr} (f) demonstrates that RadioLLM achieves superior performance at high SNR levels (greater than 0 dB), surpassing all baseline models. Even at low SNRs, it maintains competitive accuracy, albeit with performance slightly lower than MCLDNN under certain conditions. Analysis of the few-shot learning scenario, as shown in Fig.~\ref{fig:acc-snr}(a), reveals that the accuracy of all models increases with the number of training samples. Among them, MCLDNN displays the most pronounced improvement as the sample size grows. Although RadioLLM achieves comparable performance to traditional models in some extreme low-sample regimes, its advantage becomes increasingly apparent as more samples are provided, ultimately demonstrating outstanding performance as the training set expands.

On the RML16B dataset, the proposed RadioLLM model maintains outstanding performance, surpassing the runner-up SemiAMC by 2.18\% in terms of OA, as shown in Table~\ref{tab:classification_results}. This result highlights the strong classification capability and the robust feature extraction capacity of the large-scale model, which benefits from both pre-training and architectural innovations. Moreover, RadioLLM achieves the highest Kappa coefficient, further indicating its superior consistency and reliability in multi-class signal classification tasks.

In terms of few-shot learning, Fig.~\ref{fig:acc-snr} (b) reveals that MCLDNN is less adaptive to limited training data and thus exhibits relatively poor performance in low-sample scenarios. However, its accuracy increases sharply as the number of training samples grows. In contrast, RadioLLM, TcssAMR, and SemiAMC exhibit only marginal differences when trained with small sample sizes, with RadioLLM holding a slight advantage. This phenomenon can be attributed to the large parameter count of RadioLLM, which makes it challenging to optimize under extremely limited data. As the sample size increases, the performance gap widens, and RadioLLM’s advantage becomes more prominent, demonstrating its scalability and capacity to leverage larger datasets effectively.

Fig.~\ref{fig:acc-snr} (g) further demonstrates that RadioLLM excels in high SNR conditions, outperforming other models by approximately 3\%. At low SNRs, its performance remains competitive and comparable to baseline methods, indicating stable robustness under noisy environments. The confusion matrix in Fig.~\ref{fig:acc-snr} (l) reflects patterns similar to those observed on RML16A, with challenges remaining in distinguishing between closely related modulation schemes such as 16-QAM and 64-QAM, as well as noise-sensitive signals like WBFM and AM-DSB. These findings suggest that, while RadioLLM is highly effective and generally robust, future work could focus on further enhancing its discrimination ability for highly similar or ambiguous classes.

On the RML16C dataset, the proposed RadioLLM model also achieves the best performance, outperforming all competing methods with an overall accuracy of 68.19\% and a Kappa coefficient of 0.6441, as shown in Table~\ref{tab:classification_results}. In the few-shot learning setting, Fig.~\ref{fig:acc-snr} (c) shows that all models benefit from increased sample size, with RadioLLM consistently maintaining a leading position as the number of training samples grows. As illustrated in Fig.~\ref{fig:acc-snr} (h), RadioLLM exhibits robust performance across a wide range of SNRs, surpassing other models particularly at higher SNRs, which confirms its resilience to noise and superior classification capability under challenging conditions. Furthermore, the confusion matrix in Fig.~\ref{fig:acc-snr} (m) demonstrates that RadioLLM achieves excellent class discrimination overall, with no severe misclassification except for notable confusion between AM-DSB and WBFM.

To further evaluate the generalization capability of the proposed model, we extended our experiments to the RML22 dataset, which shares similar characteristics with RML16A. Notably, we did not perform any pre-training on RML22, thereby providing a rigorous test of the model’s domain transferability. As shown in Table~\ref{tab:classification_results}, RadioLLM still achieves competitive performance, surpassing TcssAMR by 1.22\% in OA and achieving the highest Kappa coefficient of 0.5488 among all compared methods.
Regarding few-shot learning, Fig.~\ref{fig:acc-snr} (d) demonstrates that RadioLLM steadily outperforms other models as the number of training samples increases, maintaining a consistent advantage across different sample size regimes. This result highlights the model’s strong adaptation ability and robustness, even when trained with limited data.

As depicted in Fig.~\ref{fig:acc-snr} (i), RadioLLM maintains robust performance across a wide range of SNR levels, and clearly outperforms the baselines at higher SNRs, further confirming its resilience to noise and its capacity for generalization in unseen domains.
The confusion matrix in Fig.~\ref{fig:acc-snr} (n) shows that while RadioLLM achieves high accuracy for most classes, the main sources of misclassification are between QPSK and 8PSK, as well as between QAM16 and QAM64. 

RadioLLM also outperformed competing models on more complex datasets. On RML18A, it managed increased task complexity better than MCLDNN, while on the Wi-Fi dataset, it achieved the highest performance among all methods tested, effectively handling intricate signal patterns. As shown in Fig.~\ref{fig:acc-snr} (o), for the RML18A dataset, misclassifications among similar modulation types, such as 8PSK and BPSK, were infrequent. This highlights the effectiveness of transfer learning in distinguishing between closely related signal types. On the ADS-B dataset, while RadioLLM did not achieve the best performance, its results were still acceptable. 

We further examine RadioLLM's adaptability under varying few-shot scenarios, as depicted in Fig.~\ref{fig:acc-snr} (a)-(e). RadioLLM consistently outperforms baseline methods across various few-shot settings, particularly evident when training samples are extremely limited (e.g., 1-shot and 5-shot). For instance, on the RML16A and RML16B datasets, RadioLLM maintains a significant accuracy advantage even with fewer than 10 labeled samples, underscoring its superior generalization capabilities. This advantage becomes more pronounced with increasing sample sizes, reflecting the effectiveness of the pre-trained feature representations obtained from large-scale data.

On more challenging datasets such as RML18A, which contains a large number of modulation classes, RadioLLM achieves a substantial performance improvement, outperforming the second-best method by 4.37\% in OA and achieving the highest Kappa coefficient, as shown in Table \ref{tab:classification_results}. In the few-shot regime, as depicted in Fig.~\ref{fig:acc-snr} (e), RadioLLM maintains a consistent advantage of approximately 2\% over the second-best model across varying sample sizes, demonstrating robust adaptation even with limited training data. Furthermore, as illustrated in Fig.~\ref{fig:acc-snr} (j), RadioLLM exhibits significant gains in high SNR conditions, achieving around 10\% higher accuracy than the next best method at SNR levels above 12~dB, highlighting its strong noise resilience.

The confusion matrix in Fig.~\ref{fig:acc-snr} (o) reveals that the primary sources of misclassification occur between 4ASK and 8ASK, 16PSK and 32PSK, as well as among the higher-order QAM classes (64QAM, 128QAM, and 256QAM).
Fig.~\ref{fig:acc-snr} (f)-(j) details the performance of RadioLLM under varying SNRs. RadioLLM consistently exhibits superior robustness to noise compared with competing methods, especially in low-SNR scenarios. Specifically, RadioLLM significantly surpasses other methods in the challenging negative SNR range (below 0 dB), where accurate modulation recognition is inherently difficult. This robustness is particularly evident in datasets such as RML16C and RML22, highlighting RadioLLM's capability to effectively extract meaningful signal features despite noisy conditions.

RadioLLM's superior performance can be attributed to its effective integration of large-scale pre-training and fine-tuning strategy. By leveraging comprehensive representations learned from extensive signal data, RadioLLM efficiently adapts to diverse modulation classification tasks, exhibiting strong transfer learning capabilities. Moreover, its sophisticated architecture, which integrates layered feature extraction mechanisms with attention-based modeling, provides meaningful hierarchical representations that effectively discriminate among subtle differences in complex modulation schemes.

In summary, our extensive evaluations clearly demonstrate RadioLLM’s exceptional classification accuracy, robust generalization across diverse datasets, and resilience under challenging noise conditions. While certain confusion remains among closely related modulation types in highly noisy contexts, these results establish RadioLLM as a highly promising method for realistic radio signal classification tasks, offering significant advantages over contemporary approaches.
\subsection{Performance Analysis of Baselines with Multi-Domain Pre-training}
To further explore the benefits of large-scale and multi-domain learning, we conducted additional experiments in which baseline methods TcssAMR and SemiAMC were subjected to pre-training on multiple datasets. Specifically, both models were first pre-trained on an aggregated dataset composed of RML16A, RML16B, and RML16C, and subsequently fine-tuned individually on each of these datasets. This experimental design aims to assess whether exposure to diverse domain distributions during pre-training can enhance the models’ generalization capacity and downstream performance.

\begin{figure}[h!]
    \centering
    \includegraphics[width=\linewidth]{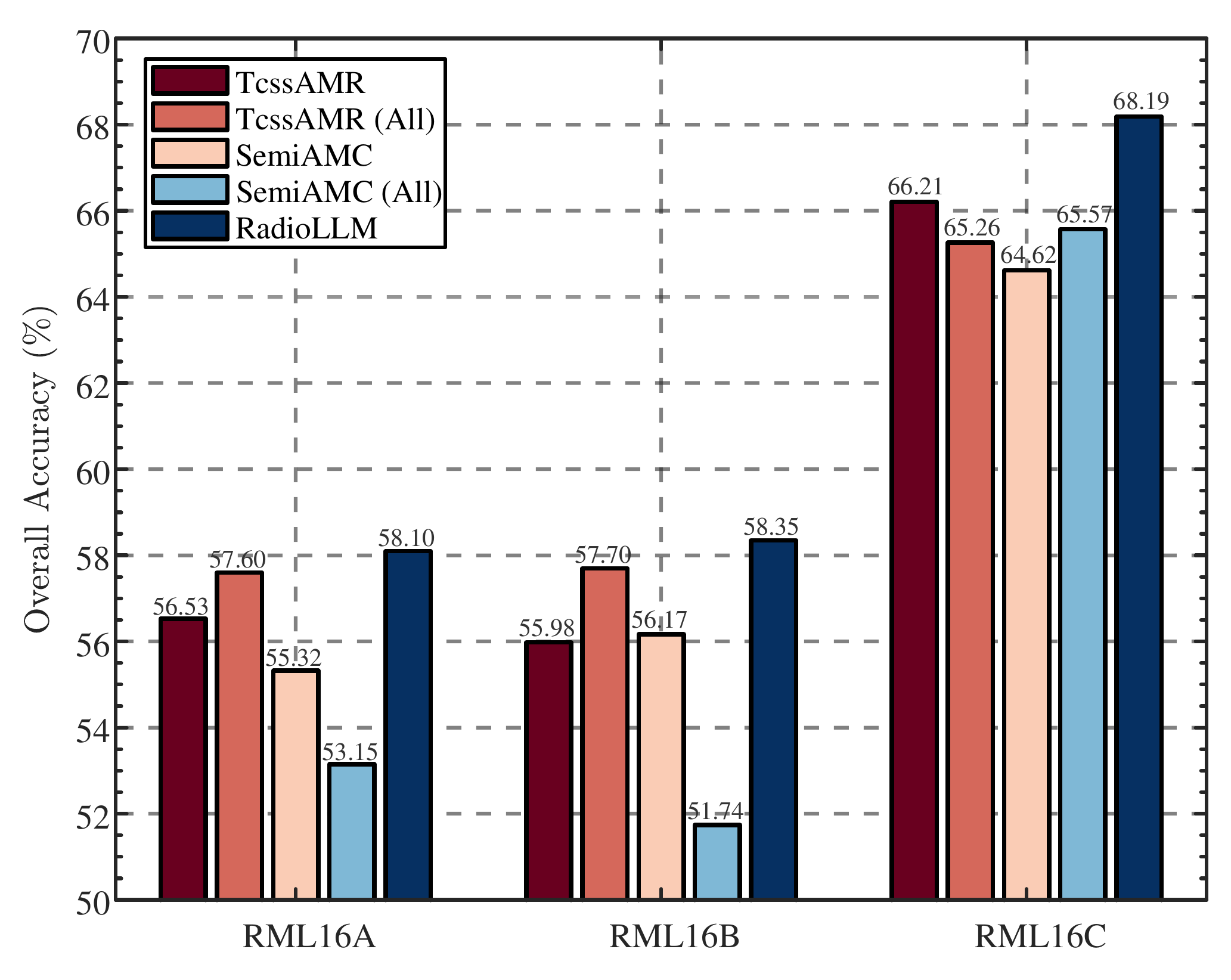}
    \caption{Comparison of OA for different methods on RML16A, RML16B, and RML16C. “All” denotes variants pre-trained jointly on all three datasets. The results illustrate the effect of multi-domain pre-training on model adaptability and generalization.}
    \label{fig:compare}
\end{figure}

The experimental results are visualized in Fig.~\ref{fig:compare}. As observed, multi-dataset pre-training generally leads to improvements across several evaluation metrics, such as OA and Kappa coefficient, compared to models trained from scratch on a single dataset. Both TcssAMR and SemiAMC exhibit higher performance on some datasets after multi-domain pre-training, indicating better feature transfer and increased robustness to domain shifts.

However, it is noteworthy that while multi-domain pre-training promotes performance gains in many scenarios, it also results in slight performance drops on certain datasets. This phenomenon is anticipated and can be attributed to the intrinsic characteristics of both the datasets and the models. The aggregated pre-training exposes the model to a broader range of data distributions, which enhances its ability to generalize across domains. Nevertheless, the same diversity may introduce domain-specific biases that are not fully aligned with the target dataset, leading to less optimal adaptation in some cases.

This result highlights a fundamental property of large-scale, pre-trained models: their adaptability to multiple datasets stems from their capacity to learn generalizable and transferable representations. Yet, the presence of both improvements and declines underscores the inherent trade-off between broad generalization and domain-specific optimization. The observed variance in performance reflects the model's bias induced by pre-training, which can either align well with or deviate from the target domain characteristics.

In summary, our findings demonstrate that large-scale, multi-domain pre-training equips baseline models with enhanced adaptability and the ability to transfer knowledge across various scenarios. Nevertheless, achieving optimal performance in real-world signal processing tasks often requires further architectural innovations or domain-specific fine-tuning to fully leverage the strengths of foundation models while mitigating the risks of domain mismatch. This underscores the importance of both large-scale pre-training for broad applicability and targeted model design for robust signal classification in complex environments.

\begin{figure*}[!t]
    \centering
    \includegraphics[width=\linewidth]{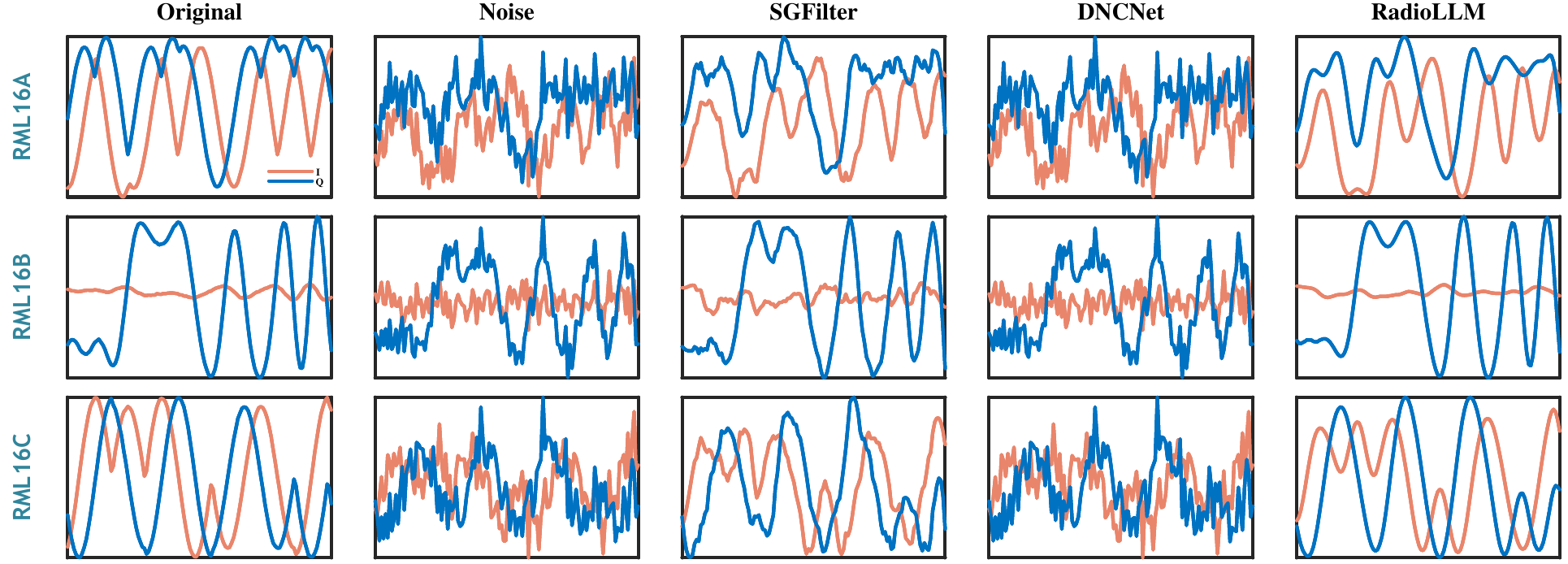}
    \caption{Comparison of denoising results on randomly selected high-SNR samples from the RML16A, RML16B, and RML16C datasets. For each dataset, the first column shows the original clean signal (Original), and the second column presents the corresponding signal after additive noise (Noise). The subsequent columns display the denoised results obtained by the SGFilter, DNCNet, and the proposed RadioLLM, respectively. RadioLLM demonstrates superior denoising performance, more effectively preserving the structure of the original signal under various conditions.}
 \label{fig:denoise}
\end{figure*}

\subsection{Comparison with Denoise methods}
%We comprehensively evaluated the proposed RadioLLM framework against both traditional signal processing and deep learning-based denoising models across three benchmark datasets: RML16A, RML16B, and RML16C. For methods lacking publicly available implementations, we re-implemented and retrained them in strict accordance with the protocols described in their original publications.
We comprehensively evaluated the proposed RadioLLM framework against both traditional signal processing and deep learning-based denoising models across three benchmark datasets: RML16A, RML16B, and RML16C. For methods lacking publicly available implementations, we re-implemented and retrained them in strict accordance with the protocols described in their original publications. During testing, to rigorously assess model robustness under realistic noise conditions, we injected additive noise into the test data, varying the SNR from 0dB to 10dB. This evaluation strategy ensures that all models are compared fairly under challenging and diverse noise environments, further highlighting the effectiveness of the proposed approach.

As illustrated in Fig.~\ref{fig:denoise}, a qualitative comparison is presented using randomly selected high-SNR samples from each dataset. The original clean signals (Original) are shown alongside their noisy counterparts (Noise), followed by the denoised results produced by SGFilter~\cite{sgfilter}, DNCNet~\cite{DNCnet}, and the proposed RadioLLM. Visually, RadioLLM is able to recover signal waveforms that are much closer to the original, effectively suppressing noise while preserving both the  I and Q components’ structures. In contrast, both SGFilter and DNCNet either oversmooth the signal or fail to adequately eliminate noise, resulting in significant distortion or residual artifacts.

\begin{table}[!h]
\centering
\caption{SSIM Values for Different Models Across Datasets.}
\label{tab:ssim-values}
\begin{tabular}{lccc}
\toprule
\textbf{Method} & \textbf{RML16A} & \textbf{RML16B} & \textbf{RML16C} \\ \midrule
SGFilter~\cite{sgfilter} & 0.782 & 0.821 & 0.777 \\
DNCNet~\cite{DNCnet}  & 0.721 & 0.742 & 0.698 \\
RadioLLM & 0.838 & 0.893 & 0.846\\
\bottomrule
\end{tabular}
\end{table}
Quantitative results, summarized in Table~\ref{tab:ssim-values}, further validate these observations. Among all evaluated methods, RadioLLM consistently achieves the highest SSIM values across all datasets (0.838, 0.893, and 0.846 for RML16A, RML16B, and RML16C, respectively), indicating superior structural preservation and fidelity in the denoised outputs. SGFilter, despite being a classical smoothing technique, achieves the second-best SSIM scores, but its fixed and non-adaptive nature limits its effectiveness in handling the complex and diverse noise patterns typical in modern wireless signals. While SGFilter performs reasonably well on simpler noise profiles, it lacks the learning capability required for more sophisticated denoising tasks.

DNCNet, a deep learning-based model, exhibits lower SSIM scores across all datasets compared to both SGFilter and RadioLLM. This suggests that, although DNCNet benefits from neural network-based modeling, it struggles to fully capture the multi-scale temporal and frequency dependencies essential for robust denoising.

In contrast, RadioLLM leverages hierarchical relational modeling and frequency-aware fusion mechanisms, enabling more effective extraction of relevant signal features and superior generalization across diverse and challenging conditions. These results, corroborated by both qualitative and quantitative analyses, highlight the robustness and adaptability of RadioLLM, firmly establishing its superiority over both traditional and deep learning-based denoising baselines.
% Compared to existing baselines, RadioLLM demonstrates clear advantages in terms of structural preservation and denoising robustness. SGFilter \cite{sgfilter}, despite being a traditional signal smoothing technique, achieves the second-highest SSIM values among the evaluated methods. However, its fixed, non-adaptive nature limits its ability to handle complex and diverse noise patterns commonly present in modern wireless signals. While SGFilter performs reasonably on simpler noise profiles, it lacks the learning capacity required for more intricate denoising tasks.
% DNCNet \cite{DNCnet}, a deep learning-based approach, shows lower SSIM performance across all datasets compared to both SGFilter and RadioLLM. This suggests that DNCNet's design, although effective to some extent, struggles to fully capture the multi-scale temporal and frequency dependencies necessary for robust denoising.
% In contrast, RadioLLM leverages hierarchical relational modeling and frequency-aware fusion, enabling superior extraction of relevant features and better generalization across diverse conditions. These results highlight the robustness and adaptability of RadioLLM, establishing its superiority over both traditional and deep learning-based denoising methods.
\subsection{Ablation Studies}
To thoroughly assess the effectiveness of the proposed HTRP and FAF modules, a series of controlled ablation experiments were conducted. These experiments systematically examined the individual and combined contributions of each module to the overall model performance, providing deeper insights into their respective roles. Unless explicitly stated otherwise, all ablation studies were performed on the RML16A dataset, following the experimental setup outlined in Section \ref{Implementation}. The results are summarized in Table \ref{tab:ablation}.

\begin{table}[htbp]
\centering
\caption{Ablation Study Results for the HTRP and FAF Modules.}
\label{tab:ablation}
\begin{tabular}{cc|cccc}
\toprule
\textbf{HTRP} & \textbf{FAF} & \textbf{OA (\%)} & \textbf{Kappa} & \textbf{SSIM} & \textbf{Inference Time(ms)}\\
\midrule
\xmark & \xmark & 55.39 & 0.5097 & 0.805 & 1131.0\\
\cmark & \xmark & 57.08 & 0.5283 & 0.833 & 766.5\\
\xmark & \cmark & 57.25 & 0.5306 & 0.857 & 1149.4\\
\cmark & \cmark & 58.10 & 0.5391 & 0.838 & 783.8\\
\bottomrule
\end{tabular}
\end{table}
As presented in Table~\ref{tab:ablation}, the baseline model, which excludes both the HTRP and FAF modules, achieves an OA of 55.39\%, a Kappa coefficient of 0.5097, a SSIM of 0.805, and an inference time of 1131.0~ms. These results indicate that, without the integration of specialized modules, the model demonstrates limited capability in both classification accuracy and structural preservation, particularly in the presence of challenging noisy conditions. Furthermore, the baseline configuration incurs a relatively high computational cost during inference.

The incorporation of the HTRP module in isolation yielded substantial and multifaceted improvements across all evaluation metrics when contrasted with the baseline configuration. As delineated in Table~\ref{tab:ablation}, the OA experienced an increase from the baseline of 55.39\% to 57.08\%. Similarly, the Kappa coefficient, a measure of agreement adjusted for chance, rose from 0.5097 to 0.5283. The SSIM, indicative of perceived image quality and structural preservation, also showed enhancement, improving from 0.805 to 0.833. This finding strongly suggests that the HTRP module not only augments the model’s capacity to effectively leverage domain-specific relational knowledge, thereby fostering the extraction of more salient temporal and relational features crucial for accurate classification, but also introduces a significant enhancement in computational performance. The pronounced reduction in inference time can be primarily attributed to the hierarchical design inherent to the HTRP module. Such an architecture is strategically conceived to streamline the propagation of information throughout the model and to minimize redundant computational steps, likely by creating more efficient pathways for feature processing or by reducing the overall parameter load associated with these operations. This gain in efficiency is of particular importance for applications deployed in real-time environments or those constrained by limited computational resources, where rapid inference and decision-making are paramount. It should be noted, however, that although the HTRP module leads to a notable increase in classification accuracy, it achieves a slightly lower SSIM compared to the FAF module. This phenomenon may be attributed to the introduction of textual information, which, while advantageous for classification, can introduce subtle conflicts with the objective of maintaining signal structure. This observation underscores an inherent trade-off between the extraction of discriminative features and optimal denoising performance.

Independent integration of the FAF module produces an OA of 57.25\%, a Kappa coefficient of 0.5306, and an SSIM of 0.857, the highest SSIM among all configurations tested. The inference time in this case is 1149.4~ms, which represents a slight increase relative to the baseline. This modest rise in computational cost is offset by the substantial gain in structural preservation and denoising quality. The FAF module’s frequency-aware fusion mechanism enables the model to capture both global and local features by effectively integrating high- and low-frequency components, thus ensuring that essential structural information is maintained even in the presence of noise. The trade-off between the minor increase in inference time and the significant improvement in SSIM underscores the practical value of the FAF module for scenarios where denoising quality is paramount.

When both HTRP and FAF modules are jointly deployed, the model achieves the best overall performance, with an OA of 58.10\%, a Kappa coefficient of 0.5391, and an SSIM of 0.838. The inference time in this configuration is 783.8~ms, which remains markedly lower than the baseline and only marginally higher than that of the HTRP-only model. This outcome demonstrates the complementary strengths of the two modules: HTRP not only boosts discriminative feature extraction and classification accuracy, but also streamlines the inference process, while FAF contributes to superior signal reconstruction at a negligible computational cost. Importantly, the combined model delivers consistent improvements across all evaluation metrics while maintaining efficient inference, highlighting its suitability for deployment in both accuracy- and latency-sensitive environments.

In summary, the ablation studies provide comprehensive evidence regarding the individual and joint contributions of the HTRP and FAF modules. The results affirm that each module addresses distinct yet complementary aspects of the overall task. The HTRP module stands out for its dual benefits of performance enhancement and significant reduction in inference time, making it highly valuable for practical applications. The FAF module, while incurring a slight increase in computational demand, delivers the highest gains in denoising quality. Their integration achieves an optimal balance between classification accuracy, structural preservation, and computational efficiency, thereby ensuring the robustness and applicability of the model in real-world, noisy environments where both performance and efficiency are critical.
\subsection{Parameter Sensitivity Analysis}
To assess the impact of key parameters on our framework's performance, we conducted a sensitivity analysis focusing on Top-K ($K$) selection, Decoder usage, and LLM choice. The experiments were performed on the RML16A dataset under a consistent 100-shot learning setting to ensure fair comparisons. The results and observations for each parameter are detailed below.
\subsubsection{Top-K}
While increasing \( K \) can improve key metrics such as OA and Kappa by capturing richer and more diverse information, it also leads to an increase in computational load, resulting in longer inference times. The Top-K selection mechanism was introduced to mitigate token redundancy by selecting the most informative tokens, thereby enhancing model efficiency without sacrificing critical information. However, as \( K \) increases, although more informative tokens are incorporated, the computational overhead during both training and inference correspondingly rises.

\begin{figure}[htbp]
    \centering
    \includegraphics[width=\linewidth]{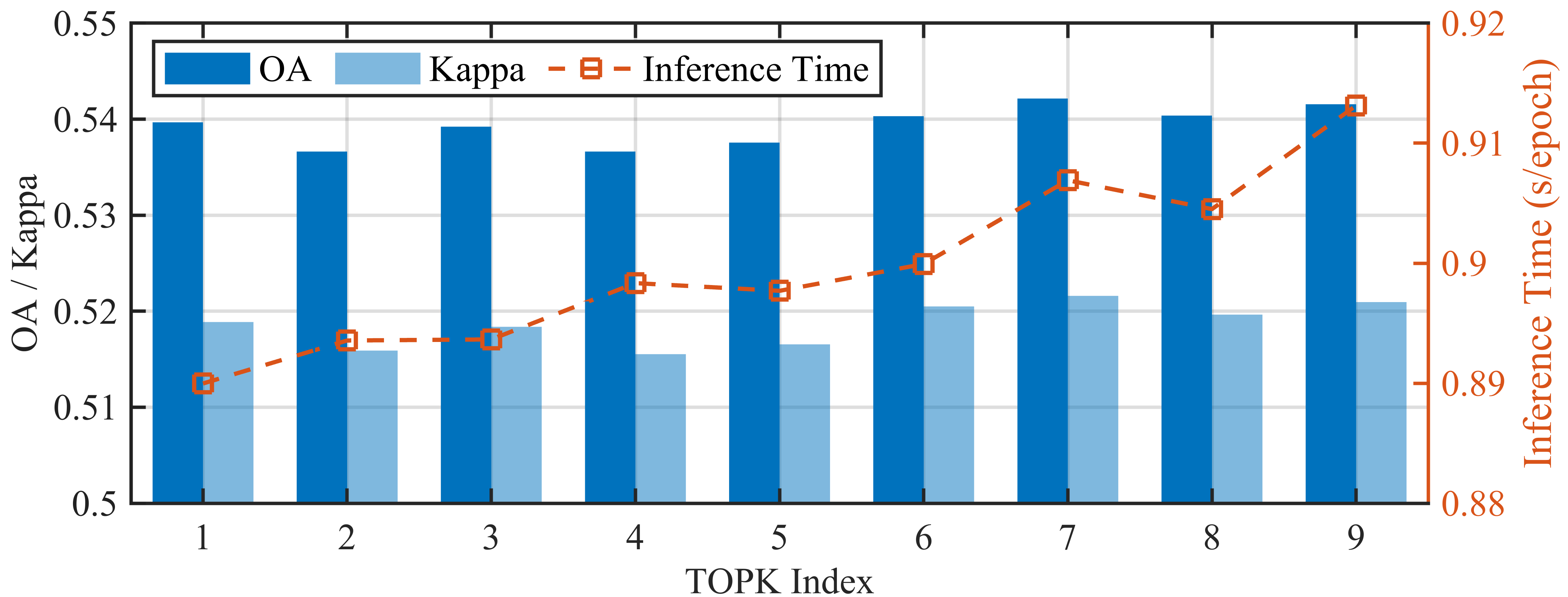}
    \caption{Performance Comparison Across Different \( K \) Values.}
    \label{fig:topk}
\end{figure}

This study systematically analyzes the sensitivity of OA, Kappa, and inference time (seconds per batch) to different values of \( K \), highlighting the inherent trade-off between classification performance and computational efficiency. As depicted in Fig.~\ref{fig:topk}, the results demonstrate a clear trend: initially, increasing \( K \) leads to notable improvements in both OA and Kappa due to the inclusion of richer contextual information. However, the inference time also increases steadily with larger \( K \) values.

Notably, \( K = 7 \) emerges as the optimal configuration, achieving the highest OA and Kappa while maintaining a reasonable inference time. Beyond this point, the marginal gains in performance diminish relative to the additional computational cost, indicating a point of diminishing returns. Therefore, careful selection of \( K \) is crucial to balance the benefits of richer token representation against the practical constraints of computational resources.
\subsubsection{Decoder Usage}
To further investigate the impact of the decoder module on the overall model performance, we conducted an ablation study in which the Transformer decoder was replaced with a Linear decoder. Experimental results indicate that substituting the Transformer decoder with a Linear decoder results in a slight yet consistent degradation in performance. Specifically, the OA decreased from 58.10\% to 57.41\%, corresponding to a relative decline of 0.69\%. Similarly, the Kappa coefficient dropped from 0.5391 to 0.5220, representing a reduction of 0.0171.

These results underscore the pivotal role of the Transformer decoder in enhancing both classification accuracy and prediction consistency. Benefiting from its global modeling capability, the Transformer decoder effectively refines contextual representations and improves information reconstruction, thereby contributing to superior task performance. In contrast, the Linear decoder lacks such global modeling capacity, leading to suboptimal results. Therefore, retaining the Transformer decoder is essential for achieving optimal performance within the proposed classification framework.
% To further examine the contribution of the decoder module to the overall model performance, we conducted an ablation study by removing the decoder and evaluating the resulting model across multiple classification tasks. The experimental results demonstrate that the absence of the decoder leads to a slight but consistent degradation in performance. Specifically, the OA decreased from 58.10\% to 57.41\%, corresponding to a relative decline of 0.69\%. Similarly, the Kappa coefficient dropped from 0.5391 to 0.5220, representing a reduction of 0.0171.

% These findings highlight the crucial role of the decoder in enhancing both classification accuracy and prediction consistency. By refining the model’s contextual understanding and improving the quality of learned representations, the decoder significantly contributes to superior task performance. Thus, maintaining the decoder component is essential for achieving optimal results in our classification framework.
\subsubsection{LLM Choice}
The choice of LLM directly influences the performance and efficiency of the proposed RadioLLM framework. To systematically evaluate the impact, we evaluated three representative LLMs: BERT \cite{bert}, GPT-2 \cite{gpt2}, and LLaMA3 \cite{llama} based on OA, Kappa coefficient, and inference time (measured in seconds per batch). The results, summarized in Table \ref{tab:llm-performance}, highlight the trade-offs between classification effectiveness and computational efficiency.

\begin{table}[htbp]
\centering
\caption{Performance Comparison Across Different LLMs.}
\label{tab:llm-performance}
\begin{tabular}{cccc}
\toprule
\textbf{LLM}       & \textbf{OA (\%)} & \textbf{Kappa} & \textbf{Inference Time (s)} \\ \midrule
BERT               & 57.53            & 0.5332         & 1.7519                       \\ 
GPT-2              & 58.10            & 0.5391         & 0.9069                       \\ 
LLaMA3             & 58.67            & 0.5480         & 0.9319                       \\ 
\bottomrule
\end{tabular}
\end{table}

As shown in Table \ref{tab:llm-performance}, LLaMA3 achieves the highest OA and Kappa coefficient among the evaluated models, indicating superior classification performance. This can be attributed to its advanced architectural design, including optimized attention mechanisms and improved parameter efficiency, as well as its training on more recent and diverse datasets. These factors collectively enhance LLaMA3’s ability to model complex patterns in radio signal data effectively.

GPT-2 demonstrates a balanced trade-off between performance and efficiency. Although its OA and Kappa are slightly lower than those of LLaMA3, it achieves the fastest inference time, making it a practical choice for latency-sensitive applications.

In contrast, BERT exhibits the lowest classification performance and the highest inference time. The increased computational cost associated with BERT can be largely attributed to its bidirectional attention mechanism, which requires attending to both past and future tokens simultaneously during each layer's computation. While this bidirectional context modeling benefits certain natural language understanding tasks, it introduces additional overhead that is less favorable for the lightweight inference requirements of radio signal classification.

Overall, these findings underscore the importance of selecting an appropriate LLM based on the specific constraints and priorities of the application, balancing accuracy and computational efficiency accordingly.
\section{Conclusion}
This paper introduces RadioLLM, an innovative framework aimed at advancing CRT by seamlessly integrating radio signal processing with LLMs. The proposed HPTR mechanism effectively bridges the gap between radio signals and LLMs by combining hardware and semantic software prompts, enabling efficient and domain-specific representations. Complementing this, the FAF module enhances the framework's ability to capture fine-grained high-frequency features, which are crucial for handling complex signal environments in CRT. Extensive experiments across diverse tasks validate RadioLLM’s effectiveness in addressing prominent challenges in CRT, such as noisy conditions, imbalanced datasets, and intricate signal patterns. These contributions establish RadioLLM as a robust and adaptable solution, paving the way for future advancements in cognitive radio systems.

% \section*{Acknowledgments}
% This should be a simple paragraph before the References to thank those individuals and institutions who have supported your work on this article.

% {\appendix[Proof of the Zonklar Equations]
% Use $\backslash${\tt{appendix}} if you have a single appendix:
% Do not use $\backslash${\tt{section}} anymore after $\backslash${\tt{appendix}}, only $\backslash${\tt{section*}}.
% If you have multiple appendixes use $\backslash${\tt{appendices}} then use $\backslash${\tt{section}} to start each appendix.
% You must declare a $\backslash${\tt{section}} before using any $\backslash${\tt{subsection}} or using $\backslash${\tt{label}} ($\backslash${\tt{appendices}} by itself
%  starts a section numbered zero.)}


% \section{References Section}
% You can use a bibliography generated by BibTeX as a .bbl file.
%  BibTeX documentation can be easily obtained at:
%  http://mirror.ctan.org/biblio/bibtex/contrib/doc/
%  The IEEEtran BibTeX style support page is:
%  http://www.michaelshell.org/tex/ieeetran/bibtex/
 
\bibliographystyle{IEEEtran} 
\bibliography{radiollm}

\vfill

\end{document}